\documentclass{PoS}
\usepackage{float}
\usepackage[force]{feynmp-auto}
\usepackage{braket}
\usepackage{slashed}
\usepackage{amsmath}
\usepackage{graphicx}
\usepackage{caption}
\usepackage{subcaption}
\newcommand{\ignore}[1]{}

\title{QED corrections to leptonic decay rates}

\ShortTitle{QED corrections to leptonic decay rates}
\author{P. A.~Boyle${}^2$, V.~G\"ulpers${}^2$, A.~J\"uttner${}^1$, C.~Lehner${}^3$, F.~\'O h\'Og\'ain${}^2$, A.~Portelli${}^2$, \newline\speaker{J. P.~Richings}${}^1$, C.T.~Sachrajda${}^1$,\\
        ${}^1$School of Physics and Astronomy, University of Southampton, Southampton SO17 1BJ, UK\\
        ${}^2$School of Physics and Astronomy, University of Edinburgh, Edinburgh EH9 3JZ, UK\\
        ${}^3$Physics Department, Brookhaven National Laboratory, Upton, NY 11973, USA\\
        E-mail: \email{j.p.richings@soton.ac.uk}}

\abstract{RBC/UKQCD is preparing a calculation of leptonic decay rates including isospin breaking corrections using a 
perturbative approach to include NLO contributions from QED effects. 
We present preliminary numerical results for a contribution to the leptonic pion decay rate and report on exploratory studies of 
computational techniques based on all-to-all propagators.}

\FullConference{The 36th Annual International Symposium on Lattice Field Theory - LATTICE2018\\
		22-28 July, 2018\\
		Michigan State University, East Lansing, Michigan, USA.}

\begin{document}

\section{Introduction}
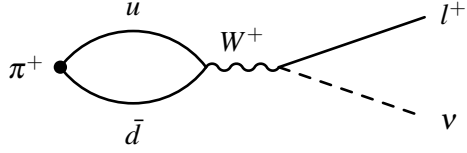
\begin{figure}[t]
	\centering
	\begin{fmffile}{pionDirectdecayW}
		\begin{fmfgraph*}(140,70)
			\fmfleft{i1}
			\fmfright{o1,o2,o3,o4,o5}		
			\fmf{plain,left=0.5,label=$u$}{i1,v1}
			\fmf{plain,left=0.5,label=$\bar{d}$}{v1,i1}
			\fmf{boson,tension=4,label=$W^+$}{v1,v2}
			\fmf{dashes}{v2,o2}
			\fmf{plain}{v2,o4}
			\fmfdot{i1}
			\fmflabel{$l^+$}{o4}
			\fmflabel{$\nu$}{o2}
			\fmflabel{$\pi^+$}{i1}
		\end{fmfgraph*}
	\end{fmffile}\vspace{-3mm}
	\caption{Pion decay into leptons via a weak current without QED contributions.}
	\label{decay0}
\end{figure}

    \noindent In this contribution we outline RBC/UKQCD's work towards a Lattice QCD calculation of the isospin-breaking (IB) corrections to the leptonic decay rate for pions and kaons. This study is motivated by the sub-percent-level precision achieved by some collaborations 
    for the calculation of $f_\pi$ and $f_K$, using various lattice actions in the isospin-symmetric limit, where up and down quarks are treated as identical particles~\cite{Aoki2017}. The aim is to determine CKM-matrix elements from leptonic decays (Figure \ref{decay0}), thus
    enabling precise tests of the Standard Model.
To date, one calculates decay constants on the lattice and then uses experimental results for the decay rates to yield the CKM matrix elements using
\vspace{-3mm}
    \begin{equation}
    \Gamma(\pi^{+} \rightarrow l^+ \nu) = \frac{m_\pi}{8\pi}G_F^2|f_{\pi^{+}}|^2|V_{ud}|^2m_{l}^2\Big(1-\frac{m_l^2}{m_\pi^2}\Big)^2 .
    \end{equation}\vspace{-5mm}
    
    \noindent The pion decay constant $f_{\pi^{+}}$ is defined in terms of the QCD matrix element, $\bra{0}\bar{d}\gamma_{\mu}\gamma_5 u\ket{\pi^{+}(p)}=ip_{\mu}f_{\pi^{+}}$, which is computed on the lattice  from Euclidean two-point correlation functions. To further  improve the precision 
    isospin-breaking (IB) effects due to the different masses of the light quarks and the difference in the QED coupling between up- and down-type quarks must be taken into account. Based on power counting in the electromagnetic coupling and the up- and down-quark mass difference, respectively,
    one expects these effects to enter at the percent level.
     
    \noindent We focus our discussion on the QED isospin-breaking corrections to leptonic decays of pions, following the approach developed in~\cite{PhysRevD.91.074506}, where the QCD+QED path integral is expanded in $\alpha$ and IR divergences are dealt with consistently.
    Naively, QED in a finite volume is ill-defined due to the appearance of photon zero modes. We subtract these by hand within 
    the framework  of QED\textsubscript{L}~\cite{Hayakawa:2008an, Borsanyi:2014jba,Blum:2010ym,Davoudi:2018qpl}. 
   %
From amongst the various possible strategies for implementing the computation we here present one based on the use of all-to-all 
propagators~\cite{Foley:2005ac}. In particular, we compute a number of low-mode eigenvectors exactly using a variant of the 
Lanczos algorithm~\cite{lanczos}. The complement of the low-mode space is then estimated stochastically. As detailed below this setup allows to 
compute contractions of quark propagators (not only the ones required for the QED corrections) off-line, i.e., without requiring a super computer.

\section{QED Isospin Breaking corrections to leptonic decay rates}
\begin{figure*}[t!]
		\centering 
				\begin{subfigure}[b]{0.25\textwidth}
					\hspace{5mm}
					\begin{fmffile}{pionDirectdecay3}
						\begin{fmfgraph*}(80,40)
							\fmfleft{i1}
							\fmfright{o1,o2}	
							\fmf{plain,left=0.5}{i1,v1}
							\fmf{plain,left=0.5}{v1,i1}
							\fmfforce{.6w,.5h}{v1}
							\fmfforce{.67w,.5h}{v2}
							\fmf{dashes}{v2,o1}
							\fmf{plain}{v2,o2}
							\fmffreeze
							\fmfforce{.3w,.8h}{v3}
							\fmfforce{0.8w,0.8h}{v4}
							\fmf{photon,left=0.5,label=$\gamma$}{v3,v4}
							\fmfdot{i1}
							\fmfdot{v1}
							\fmfdot{v2}
							\fmfdot{v3}
							\fmfdot{v4}
							\fmfv{label=$l$,label.angle=1}{o2}
							\fmfv{label=$\nu$,label.angle=1}{o1}
							\fmflabel{$\pi / k$}{i1}
						\end{fmfgraph*}
					\end{fmffile}
					\caption{\quad\quad\quad}
					\label{lcd}
				\end{subfigure}%
				\hspace{-1mm}
				\begin{subfigure}[b]{0.25\textwidth}
					\hspace{5mm}
					\begin{fmffile}{pionDirectdecayexch}
						\begin{fmfgraph*}(80,40)
							\fmfleft{i1}
							\fmfright{o1,o2}	
							\fmf{plain,left=0.5}{i1,v1}
							\fmf{plain,left=0.5}{v1,i1}
							\fmfforce{.6w,.5h}{v1}
							\fmfforce{.67w,.5h}{v2}
							\fmf{dashes}{v2,o1}
							\fmf{plain}{v2,o2}
							\fmffreeze
							\fmfforce{.3w,.8h}{v3}
							\fmfforce{.3w,.2h}{v4}
							\fmf{photon}{v3,v4}
							\fmfdot{i1}
							\fmfdot{v1}
							\fmfdot{v2}
							\fmfdot{v3}
							\fmfdot{v4}
						\fmfv{label=$l$,label.angle=1}{o2}
						\fmfv{label=$\nu$,label.angle=1}{o1}
							\fmflabel{$\pi / k$}{i1}
						\end{fmfgraph*}
					\end{fmffile}
					\caption{\quad\quad\quad}
					\label{ecd}
				\end{subfigure}%
				\hspace{-1mm}
				\begin{subfigure}[b]{0.25\textwidth}
				    \hspace{5mm}
					\begin{fmffile}{pionDirectdecayself}
						\begin{fmfgraph*}(80,40)
							\fmfleft{i1}
							\fmfright{o1,o2}	
							\fmf{plain,left=0.5}{i1,v1}
							\fmf{plain,left=0.5}{v1,i1}
							\fmfforce{.6w,.5h}{v1}
							\fmfforce{.67w,.5h}{v2}
							\fmf{dashes}{v2,o1}
							\fmf{plain}{v2,o2}
							\fmffreeze
							\fmfforce{.1w,.7h}{v3}
							\fmfforce{.5w,.7h}{v4}
							\fmf{photon,left=1}{v3,v4}
							\fmfdot{i1}
							\fmfdot{v1}
							\fmfdot{v2}
							\fmfdot{v3}
							\fmfdot{v4}
						\fmfv{label=$l$,label.angle=1}{o2}
						\fmfv{label=$\nu$,label.angle=1}{o1}
							\fmflabel{$\pi / k$}{i1}
						\end{fmfgraph*}
					\end{fmffile}
					\caption{\quad\quad\quad}
					\label{sed}
				\end{subfigure}%
				\hspace{-1mm}
				\begin{subfigure}[b]{0.25\textwidth}
					\hspace{5mm}
					\begin{fmffile}{pionDirectdecaytadpole}
						\begin{fmfgraph*}(80,40)
							\fmfleft{i1}
							\fmfright{o1,o2}	
							\fmf{plain,left=0.5}{i1,v1}
							\fmf{plain,left=0.5}{v1,i1}
							\fmfforce{.6w,.5h}{v1}
							\fmfforce{.67w,.5h}{v2}
							\fmf{dashes}{v2,o1}
							\fmf{plain}{v2,o2}
							\fmffreeze
							\fmfforce{.3w,.6h}{v3}
							\fmfforce{.3w,.2h}{v4}
							\fmf{photon,left=1}{v3,v4}
							\fmf{photon,left=1}{v4,v3}
							\fmfdot{i1}
							\fmfdot{v1}
							\fmfdot{v2}
							\fmfdot{v4}
						    \fmfv{label=$l$,label.angle=1}{o2}
						    \fmfv{label=$\nu$,label.angle=1}{o1}
							\fmflabel{$\pi / k$}{i1}
						\end{fmfgraph*}
					\end{fmffile}
					\caption{\quad\quad\quad}
	            	\label{tpd}					
				\end{subfigure}\vspace{-3mm}
	\caption{\footnotesize The four connected contributions to the QED-isospin-breaking correction to leptonic decay of a meson without final-state photon. The perturbative expansion contains the (a) lepton coupling, (b) exchange, (c) self-energy and (d) tadpole diagrams.}\label{QEDexp}\vspace{-0.4cm}
	\end{figure*}
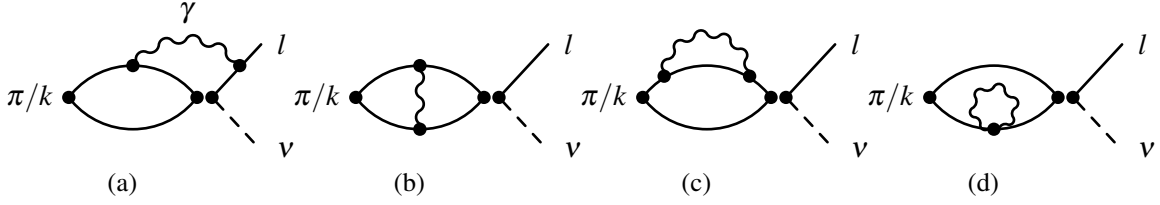

\noindent In order to calculate the infra-red (IR) finite order-$\alpha$ leptonic decay rate, we must consider contributions from graphs with and without final state photons to cancel IR divergences~\cite{Bloch:1937pw}. Here we follow the strategy outlined in~\cite{PhysRevD.91.074506} to carefully deal with IR divergences, where the contributions with final state photons are treated analytically using the point-like approximation. 
The lattice computation therefore only has to deal with diagrams without final-state photons for which the perturbative expansion of the
QCD+QED path integral~\cite{deDivitiis:2013xla} to order $\alpha$ generates the set of diagrams illustrated in figure
\ref{QEDexp}.
\noindent We use Feynman gauge for which 
the photon propagator takes the form,
		\begin{equation}
			\Delta_{\mu\nu} (x-y) = \delta_{\mu\nu}\frac{1}{L^3T}\sum_{k,\vec{k}\ne 0}\frac{e^{ik.(x-y)}}{\hat{k}^2} \, =  \braket{A_{\mu}(x)A_{\nu}(y)},
		\end{equation}
where $\hat{k}=\frac a2\sin(\frac {ak}2)$ is the lattice momentum of the photon and $L$ and $T$ the spatial and time extent of the lattice, respectively. In practice, the photon propagator is generated by inserting stochastic photons~\cite{Giusti:2017dmp}. Figure \ref{QEDexp} shows illustrations of
connected contributions without final-state photon. Figure \ref{lcd} is a contribution where the photon couples to a quark and a lepton. We implement the lepton propagator on the lattice as a free domain wall fermion. As a first test the we calculated diagram in Figure \ref{lcd} using sequential propagators~\cite{Boyle:2008rh} on a $24^3 \times 64$ lattice. The ensemble used has an isospin symmetric pion mass of $340$~MeV and inverse lattice spacing of $a^{-1} = 1.78$ GeV~\cite{Allton:2008pn}.
    \begin{figure}
\centering
			\includegraphics[width=0.6\textwidth]{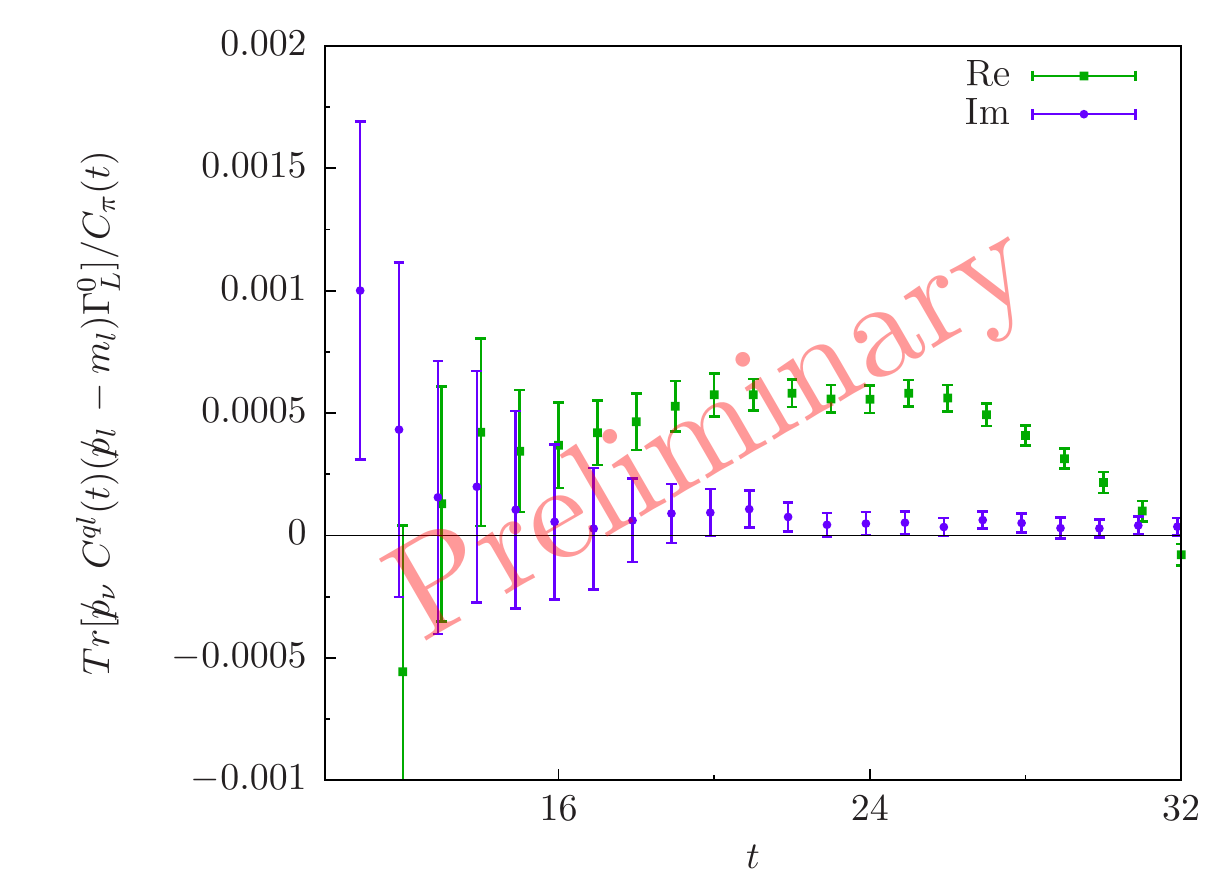}
    		\flushleft
    		\caption{The plot shows the real (green) and imaginary (blue) components of the correlator for the diagram in figure \ref{lcd} when the leptonic trace is formed. The horizontal axis is the time difference between the weak operator and the meson source positions. The vertical axis is the correlator for the lepton coupling diagram $C^{ql}$ with the trace over the free spin indices completed by including the factor $(\slashed{p}_l -m_l)\Gamma_L^0\slashed{p}_\nu$, normalised by the pion correlator.
    			}\label{graph}
    \end{figure}

    \noindent The results for diagram \ref{lcd} illustrated in Figure \ref{graph} shows an encouraging signal but we also investigate other methods which might offer better signal.
    \vspace{-4mm}
\section{The all-to-all approach and meson fields}
    \noindent We follow the all-to-all approach in~\cite{Foley:2005ac} where the propagator is decomposed into a number of exact low-mode eigenvectors the complement which is solved  stochastically. We now describe how this offers a  convenient way of structuring the calculation of correlators. 
Following~\cite{Foley:2005ac} the all-to-all propagator can be constructed from two sets of vectors, $v_i(x)$ and $w_i(x)$, such that
    \begin{equation}\label{a2aprop}
    D_\textrm{A2A}^{-1}(x,y) = \sum_{i=1}^{N_\textrm{modes}} v_i^{ }(x)w_i^{\dagger}(y)\,,
    \end{equation}
    where     $v_l(x) = \phi_l(x)$ and $w_l(y) = \phi_l(y)/\lambda_l$ are exact eigenvectors of the Dirac operator
    $\phi_l(x)$ with eigenvalues $\lambda_l$.
   In practice we only ever compute a limited number $N_l$. We then  estimate their complement stochastically. In particular, 
    we write
    \begin{equation}
    D_\textrm{A2A}^{-1}(x,y) = \sum_{l=1}^{N_l} v_l^{ }(x)w_l^{\dagger}(y) + \sum_{h=N_l+1}^{N_\textrm{modes}} v_h^{ }(x)w_h^{\dagger}(y)\,,
    \end{equation}
    where $N_\textrm{modes} = N_\textrm{high} + N_\textrm{low}$.
%
To calculate the high modes we use stochastic noise sources $\eta_h=\{\pm 1\}+i\{\pm 1\}=w_h$ from which we project out the
 low mode contribution to the propagator,
    \vspace{-3mm}
    \begin{equation}
    v_h(x) = \ignore{D_\textrm{defl}^{-1}\eta_{h}(x) =} \big( D^{-1} - \sum_{l=1}^{N_l}\phi_l^{ }(x)\phi_l^{\dagger}(x)/\lambda_l\big)\eta_{h}(x).
    \end{equation}
    \subsection{Two point correlation function}

    \noindent We can consider a two point correlation function and rewrite it in terms of all-to-all propagators using (\ref{a2aprop}),

    \begin{figure}
    	\centering
    	\begin{fmffile}{pionDirect}
    		\begin{fmfgraph*}(80,40)
    			\fmfleft{i1}
    			\fmfright{o1}
    			\fmftop{v1}
    			\fmfbottom{v2}
    			\fmf{plain,left=0.25,fore=blue,label=$v_i(x)$}{i1,v1}
    			\fmf{plain,left=0.25,fore=red,label=$w_i^{\dagger}(y)$}{v1,o1}
    			\fmf{plain,left=0.25,fore=blue,label=$w_j^{\dagger}(x)$}{v2,i1}
    			\fmf{plain,left=0.25,fore=red,label=$v_j(y)$}{o1,v2}
    			\fmfdot{i1}
    			\fmfdot{o1}
    			\fmflabel{$\Gamma_1$}{i1}
    			\fmflabel{$\Gamma_2$}{o1}
    		\end{fmfgraph*}
    	\end{fmffile}
    	\vspace{3mm}
		 \caption{A two point function.}
    \end{figure}
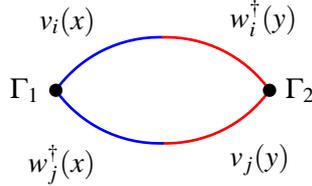
    \vspace{-7mm}
	\begin{align}
  	C_{\Gamma_1\Gamma_2}(t_y-t_x) = &\sum_{\vec{x},\vec{y}}Tr[\Gamma_1 S(x,y) \Gamma_2 S(y,x)]
	 =\sum_{\vec{x},\vec{y}}tr\left[\Gamma_1\sum_{i=1}^{N_\textrm{modes}}v^{ }_i(x)w_i^{\dagger}(y)\Gamma_2\sum_{j=1}^{N_\textrm{modes}}v^{ }_j(y)w_j^{\dagger}(x)\right] \\
	= &\, \sum_{i,j=1}^{N_\textrm{modes}}tr\left[\Pi_{ij}(t_x;\Gamma)\Pi_{ij}(t_y;\Gamma)   \right]\label{MF2pt}\,,\nonumber
	\end{align}
where we have introduced the shorthand 
    \begin{equation}\label{MF}
    \Pi_{ij}(t_x;\Gamma) = \sum_{\vec{x}}w_i^{\dagger}(x)\Gamma v_j(x)\,,
    \end{equation}
    which we refer to colloquially as \emph{meson field}. Here, $\Gamma$ represents any choice of gamma matrix.  Meson fields $\Pi_{ij}$ are of size $N_{T}\times N^2_{modes}$, where $N_T$ is the time extent of the lattice. The spatially summed meson fields can be stored to disk and retrieved 
    later, i.e., off-line, to compute traces of products of it. By including a phase factor the meson fields can be projected on any desired lattice momentum. 
    In this way we can construct, for instance, the  two-point function between pseudo-scalar and axial-vector currents,
        \begin{equation}
    C_{PA}(t_y-t_x) = \sum_{i,j} tr \Big[\Pi_{ij}(t_x;\gamma_5) \Pi_{ji}(t_y;\gamma_0\gamma_5)\Big]\,,
    \end{equation}
from which the lepton decay constant is determined.
\section{Meson fields and isospin-breaking corrections to leptonic decays}
%
Here we discuss the construction of the meson fields that are required for computing the QED-isospin corrections to leptonic decay. In particular, we 
consider the quark-photon coupling via the conserved vector current. 
\subsection{Point-split-operator meson fields}
For simplicity we consider the case of the conserved current for Wilson fermions,
	\begin{equation}\label{conserved2}
		V_{\mu}^{c}(x) = \frac{1}{2}\big[\bar{\psi}(x+\hat{\mu})(1+\gamma_\mu)U_{\mu}^{\dagger}(x)\psi(x) - \bar{\psi}(x)(1-\gamma_\mu)U_{\mu}(x)\,\psi(x+\hat{\mu})\big].
	\end{equation}
The structure of Eq.~(\ref{conserved2}) suggests that
we require meson fields for operators with gauge-invariant point-split structure.
	The exchange diagram as illustrated in Fig.~\ref{ecd}, with photons inserted using conserved vector currents, has the form,
	\begin{equation}
	C^{\rm exch.}(t_y-t_x) = \sum_{\vec{x},\vec{y},r,s}\braket{\bar\psi(y)\Gamma_1\psi(y) \,V_{\mu}^{c}(z_1) \,\bar\psi(x)\Gamma_2\psi(x) \,V_{\nu}^{c}(z_2)\,\Delta_{\mu\nu}(x-y)} \,.
	\end{equation}
We now rewrite the conserved-current contribution to this correlator in terms of a meson field. Concentrating on the second term in Eq.~(\ref{conserved2}) we write
    \begin{align}
    \sum_{\vec{z_1}} \braket{...\;\Gamma_1\psi(y) [\bar{\psi}(z_1)(1-\gamma_\mu)U_\mu(z_1)\psi(y+\hat{\mu})] \bar{\psi}(x)\Gamma_2\;...}\qquad\qquad\\\nonumber
    \qquad\qquad\qquad\qquad= \sum_{\vec{z_1}} \braket{...\;\Gamma_1S(y,z_1)(1-\gamma_\mu)U_\mu(z_1)S(z_1+\hat{\mu},x)\Gamma_2\;...},
    \end{align} 
   where we have carried out the Wick contractions in the 2nd line. Rewriting the propagators in terms of the all-to-all decomposition (\ref{a2aprop}),
    \begin{equation}\label{45}
    \sum_{ij} \braket{...\;v_{i}(y)\Big[\sum_{\vec{z_1}}w_{i}^{\dagger}(z_1)(1-\gamma_\mu)U_\mu(z_1)v_{j}(z_1+\hat{\mu})\Big]w_{j}^{\dagger}(x)\;...}\,.
    \end{equation}
    The square brackets on the right of (\ref{45}) contain  a meson field. We can treat the first term in (\ref{conserved2}) in the  same way 
    and thereby obtain a meson field for the conserved vector current.
    Including a stochastic photon field $A_\mu(x)$ is straight-forward,
    \begin{equation}
	    \Pi_{ij}\Big[t_x, V^c_\mu A_\mu\Big]=\sum_{\vec{x},\mu}\frac{1}{2}\Big[w_i^{\dagger}(x+\hat{\mu})(1+\gamma_\mu)U_{\mu}^{\dagger}(x)A_\mu(x)v_j(x)- w_i^{\dagger}(x)(1-\gamma_\mu)U_{\mu}(x)A_\mu(x)v_j(x+\hat{\mu})\Big].
    \end{equation}
    and allows us to construct the diagrams in Fig.~\ref{QEDexp} in terms of traces over products of meson fields.
    This method for dealing with conserved currents in the all-to-all set-up has also been understood for the DWF and overlap cases.
    
    In practice we generate sets of meson fields for different $\gamma$-structures and with and without photon fields on a super computer.
    The contraction of meson fields to form correlation functions can be done off-line on a single node, increasing flexibility.

A further meson field is required for the lepton coupling diagram in Figure \ref{lcd}. We place the leptonic part $L = \Gamma^\mu_W D^{-1}_\textrm{lepton} V_\mu^c A_\mu$ with the left-handed V-A current $\Gamma^\mu_W = \gamma_\mu(1-\gamma_5)$ in the meson field for the decay operator.
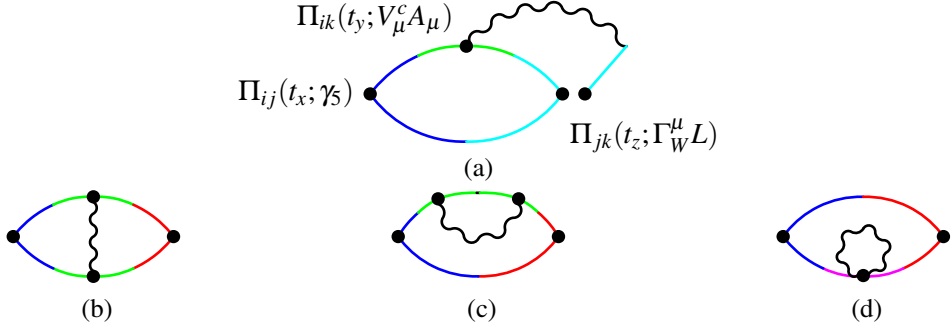
\begin{figure}
   			\begin{subfigure}[b]{\textwidth}
   				\hspace{58mm}
   				\begin{fmffile}{pionDirectdecay1}
   					\begin{fmfgraph*}(120,60)
   						\fmfleft{i1}
   						\fmfright{o1,o2}
   						\fmfforce{.3w,.8h}{v4}
   						\fmfforce{.15w,.74h}{v5}
   						\fmfforce{.45w,.74h}{v6}
   						\fmfforce{.3w,.2h}{v7}
   						\fmf{plain,fore=blue,left=0.125}{i1,v5}
   						\fmf{plain,fore=green,left=0.125}{v5,v4}
   						\fmf{plain,fore=green,left=0.125}{v4,v6}
   						\fmf{plain,fore=green+blue,left=0.125}{v6,v1}	
   						\fmf{plain,fore=green+blue,left=0.25}{v1,v7}
   						\fmf{plain,fore=blue,left=0.25}{v7,i1}
   						\fmfforce{.6w,.5h}{v1}
   						\fmfforce{.67w,.5h}{v2}
   						\fmf{plain,fore=green+blue}{v2,v3}
   						\fmffreeze
   						\fmftop{v3,v4}
   						\fmfforce{.3w,.8h}{v4}
   						\fmfforce{0.8w,0.8h}{v3}
   						\fmf{phantom}{v2,v3,o2}
   						\fmf{photon,left=0.5}{v4,v3}
   						\fmfdot{i1}
   						\fmfdot{v1}
   						\fmfdot{v2}
   						\fmfdot{v4}
   						\fmfforce{.58w,.4h}{v8}
   						\fmflabel{$\Pi_{ij}(t_x;\gamma_5)$}{i1}
   						\fmflabel{$\Pi_{ik}(t_y;V^c_\mu A_\mu)$}{v4}
   						\fmflabel{$\Pi_{jk}(t_z;\Gamma_W^\mu L)$}{v8}
   					\end{fmfgraph*}
   				\end{fmffile}
   				\vspace{-5mm}
   				\caption{\vspace{1mm}}
   				\label{clcd}
   			\end{subfigure}
			\begin{subfigure}[b]{0.33\textwidth}
				\centering
				\begin{fmffile}{pionDirectphotonexchangecolour}
					\begin{fmfgraph*}(60,30)
						\fmfleft{i1}
						\fmfright{o1}
						\fmftop{v1,v2,v3,v4,v5}
						\fmfbottom{v6,v7,v8,v9,v10}
						\fmfforce{.25w,.9h}{v11}
						\fmfforce{.75w,.9h}{v12}
						\fmfforce{.25w,.1h}{v13}
						\fmfforce{.75w,.1h}{v14}
						\fmf{plain,left=0.125,fore=blue}{i1,v11}
						\fmf{plain,left=0.125,fore=green}{v11,v3}
						\fmf{plain,left=0.125,fore=green}{v3,v12}
						\fmf{plain,left=0.125,fore=red}{v12,o1}
						\fmf{plain,left=0.125,fore=red}{o1,v14}
						\fmf{plain,left=0.125,fore=green}{v14,v8}
						\fmf{plain,left=0.125,fore=green}{v8,v13}
						\fmf{plain,left=0.125,fore=blue}{v13,i1}
						\fmf{photon,tension=2}{v3,v8}
						\fmfdot{i1}
						\fmfdot{o1}
						\fmfdot{v3}
						\fmfdot{v8}
					\end{fmfgraph*}
				\end{fmffile}
		         \caption{}
		         \label{cexch}
			\end{subfigure}
			\begin{subfigure}[b]{0.33\textwidth}
				\centering
				\begin{fmffile}{pionDirectphotonselfcolour}
					\begin{fmfgraph*}(60,30)
						\fmfleft{i1}
						\fmfright{o1}
						\fmftop{v1,v2,v3,v4,v5}
						\fmfbottom{v6}
						\fmfforce{.125w,.8h}{v11}
						\fmfforce{.875w,.8h}{v12}
						\fmfforce{.49w,1.05h}{v13}
						\fmfforce{.51w,1.05h}{v14}
						\fmf{plain,left=0.08333,fore=blue}{i1,v11}
						\fmf{plain,left=0.08333,fore=green}{v11,v2}
						\fmf{plain,left=0.1,fore=green}{v2,v13}
						\fmf{plain}{v13,v14}						
						\fmf{plain,left=0.1,fore=green}{v14,v4}
						\fmf{plain,left=0.08333,fore=green}{v4,v12}
						\fmf{plain,left=0.08333,fore=red}{v12,o1}
						\fmf{plain,left=0.25,fore=red}{o1,v6}
						\fmf{plain,left=0.25,fore=blue}{v6,i1}
						\fmf{photon,left=-1, tension=0.5}{v2,v4}
						\fmfdot{i1}
						\fmfdot{o1}
						\fmfdot{v2}
						\fmfdot{v4}
					\end{fmfgraph*}
				\end{fmffile}
				\caption{}
				\label{cself}
			\end{subfigure}
			\begin{subfigure}[b]{0.33\textwidth}
				\centering
				\begin{fmffile}{pionDirectphotontadpolecolour}
					\begin{fmfgraph*}(60,30)
						\fmfleft{i1}
						\fmfright{o1}
						\fmftop{v1,v2,v3}
						\fmfbottom{v4,v5,v6,v7,v8}
						\fmfforce{.25w,.1h}{v11}
						\fmfforce{.75w,.1h}{v12}
						\fmf{plain,left=0.25,fore=blue}{i1,v2}
						\fmf{plain,right=0.25,fore=red}{o1,v2}
						\fmf{plain,left=0.125,fore=red}{o1,v12}
						\fmf{plain,left=0.125,foreground=red+blue}{v12,v6}
						\fmf{plain,left=0.125,fore=red+blue}{v6,v11}
						\fmf{plain,left=0.125,fore=blue}{v11,i1}
						\fmffreeze
						\fmfforce{.5w,.6h}{v8}
						\fmfforce{.5w,.01h}{v9}
						\fmf{photon,tension=1,left=1}{v9,v8}
						\fmf{photon,tension=1,left=1}{v8,v9}
						\fmfdot{i1}
						\fmfdot{o1}
						\fmfdot{v9}
					\end{fmfgraph*}
				\end{fmffile}
				\caption{}
				\label{ctadp}
			\end{subfigure}
   		\caption{The diagrams shown here correspond to the QED correction to the decay rate. Here the diagrams are coloured corresponding to the type of meson field used to construct the correlator.}
\end{figure}	
    \subsection{Leptonic decay corrections from meson fields}
    \noindent Using the meson fields discussed above it is possible to construct all the diagrams required for a calculation of corrections to the decay rate. This is illustrated in Figures \ref{clcd}-\ref{ctadp} where the colours correspond to the different meson fields required to construct each graph. In particular, blue for $\gamma_5$, red for $\gamma_0\gamma_5$, green for the conserved vector current with a photon insertion, light blue for the weak Hamiltonian and lepton insertion and pink for the tadpole insertion. In total five meson fields are required to determine the QED IB corrections to the decay rate.
    
The quark-disconnected diagrams (Figure \ref{discon}) can be formed from the same set of meson fields without any further inversions.

\begin{figure}[H]
	\hspace{1cm}
\begin{minipage}{0.5\textwidth}
	\begin{figure}[H]
		\centering
		\begin{fmffile}{Disconected1}
			\begin{fmfgraph*}(80,40)
				\fmfleft{i1}
				\fmfright{o1,o2}	
				\fmf{plain,left=0.5}{i1,v1}
				\fmf{plain,left=0.5}{v1,i1}
				\fmfforce{.6w,.5h}{v1}
				\fmfforce{.67w,.5h}{v2}
				\fmf{dashes}{v2,o1}
				\fmf{plain}{v2,o2}
				\fmffreeze
				\fmftop{v3,v4}
				\fmfforce{.8w,.8h}{v3}
				\fmfforce{.6w,1h}{v4}
				\fmfforce{.4w,1h}{v5}
				\fmf{phantom,left=0.25}{i1,v4,v1}
				\fmf{phantom}{v2,v3,o2}
				\fmf{photon,left=0.5}{v4,v3}
				\fmf{plain,left=1,fore=green}{v4,v5}
				\fmf{plain,left=1,fore=green}{v5,v4}
				\fmfdot{i1}
				\fmfdot{v1}
				\fmfdot{v2}
				\fmfdot{v3}
				\fmfdot{v4}
			\end{fmfgraph*}
		\end{fmffile}
	\end{figure}
\end{minipage}
\hspace{-2cm}
\begin{minipage}{0.5\textwidth}
	\begin{figure}[H]
		\centering
		\begin{fmffile}{Disconected2}
			\begin{fmfgraph*}(80,40)
				\fmfleft{i1}
				\fmfright{o1,o2}	
				\fmf{plain,left=0.5}{i1,v1}
				\fmf{plain,left=0.5}{v1,i1}
				\fmfforce{.6w,.5h}{v1}
				\fmfforce{.67w,.5h}{v2}
				\fmf{dashes}{v2,o1}
				\fmf{plain}{v2,o2}
				\fmffreeze
				\fmftop{v3,v4}
				\fmfforce{.3w,.8h}{v3}
				\fmfforce{-.1w,1h}{v4}
				\fmfforce{.1w,1h}{v5}
				\fmf{phantom,left=0.25}{i1,v4,v1}
				\fmf{phantom}{v2,v3,o2}
				\fmf{photon,left=0.5}{v5,v3}
				\fmf{plain,left=1,fore=green}{v4,v5}
				\fmf{plain,left=1,fore=green}{v5,v4}
				\fmfdot{i1}
				\fmfdot{v1}
				\fmfdot{v2}
				\fmfdot{v3}
				\fmfdot{v5}
			\end{fmfgraph*}
		\end{fmffile}
	\end{figure}
\end{minipage}

\begin{minipage}{0.33\textwidth}
	\begin{figure}[H]
		\centering
		\begin{fmffile}{Disconected5}
			\begin{fmfgraph*}(80,40)
				\fmfleft{i1}
				\fmfright{o1,o2}	
				\fmf{plain,left=0.5}{i1,v1}
				\fmf{plain,left=0.5}{v1,i1}
				\fmfforce{.6w,.5h}{v1}
				\fmfforce{.67w,.5h}{v2}
				\fmf{dashes}{v2,o1}
				\fmf{plain}{v2,o2}
				\fmffreeze
				\fmfforce{.3w,.8h}{v3}
				\fmfforce{.7w,1h}{v4}
				\fmfforce{.5w,1h}{v5}
				\fmfforce{.0w,1h}{v6}
				\fmfforce{.2w,1h}{v7}
				\fmf{photon,left=0.5}{v7,v5}
				\fmf{plain,left=1,fore=green}{v4,v5}
				\fmf{plain,left=1,fore=green}{v5,v4}
				\fmf{plain,left=1,fore=green}{v6,v7}
				\fmf{plain,left=1,fore=green}{v7,v6}
				\fmfdot{i1}
				\fmfdot{v1}
				\fmfdot{v2}
				\fmfdot{v5}
				\fmfdot{v7}
			\end{fmfgraph*}
		\end{fmffile}
	\end{figure}
\end{minipage}
\begin{minipage}{0.33\textwidth}
	\begin{figure}[H]
		\centering
		\begin{fmffile}{Disconected3}
			\begin{fmfgraph*}(80,40)
				\fmfleft{i1}
				\fmfright{o1,o2}	
				\fmf{plain,left=0.5}{i1,v1}
				\fmf{plain,left=0.5}{v1,i1}
				\fmfforce{.6w,.5h}{v1}
				\fmfforce{.67w,.5h}{v2}
				\fmf{dashes}{v2,o1}
				\fmf{plain}{v2,o2}
				\fmffreeze
				\fmfforce{.3w,.8h}{v3}
				\fmfforce{.7w,1h}{v4}
				\fmfforce{.5w,1h}{v5}
				\fmf{photon}{v5,v4}
				\fmf{plain,left=1,fore=green}{v4,v5}
				\fmf{plain,left=1,fore=green}{v5,v4}
				\fmfdot{i1}
				\fmfdot{v1}
				\fmfdot{v2}
				\fmfdot{v4}
				\fmfdot{v5}
			\end{fmfgraph*}
		\end{fmffile}
	\end{figure}
\end{minipage}%
\begin{minipage}{0.33\textwidth}
	\begin{figure}[H]
		\centering
		\begin{fmffile}{Disconected4}
			\begin{fmfgraph*}(80,40)
				\fmfleft{i1}
				\fmfright{o1,o2}	
				\fmf{plain,left=0.5}{i1,v1}
				\fmf{plain,left=0.5}{v1,i1}
				\fmfforce{.6w,.5h}{v1}
				\fmfforce{.67w,.5h}{v2}
				\fmf{dashes}{v2,o1}
				\fmf{plain}{v2,o2}
				\fmffreeze
				\fmfforce{.3w,.8h}{v3}
				\fmfforce{.65w,1h}{v4}
				\fmfforce{.5w,1h}{v5}
				\fmfforce{.35w,1h}{v6}
				\fmf{plain,right=1,fore=red+blue}{v5,v6}
				\fmf{plain,right=1,fore=red+blue}{v6,v5}
				\fmf{photon,left=1}{v4,v5}
				\fmf{photon,left=1}{v5,v4}
				\fmfdot{i1}
				\fmfdot{v1}
				\fmfdot{v2}
				\fmfdot{v5}
			\end{fmfgraph*}
		\end{fmffile}
	\end{figure}
\end{minipage}
\caption{The set of disconnected diagrams that contribute to the QED correction to a leptonic decay at order $\alpha$. The colour coding corresponds to the meson fields that are used to construct the disconnected part.}
\label{discon}
\end{figure}
\vspace{-7mm}

\section{Conclusion}

\noindent Progress is being made towards a determination of the isospin breaking corrections to leptonic decays of pions and kaons. Here we presented one way of organising the workflow, namely in terms of meson fields, which offer a convenient approach for computing $n$-point functions 
from simple building blocks off-line. Apart from a few technicalities the above discussion carries over to Domain Wall Fermions, which we 
use in our core simulation program. 
We are in the process of testing the all-to-all approach and implementation of meson-field generation. Once we verify the all-to-all method for the calculation of QED effects we aim to calculate IB correction to leptonic decays for both the pion and kaon in this way. If this approach is successful a number of physics processes can be calculated from a set of stored meson fields, increasing the physics output from consumed computer time. 

\section*{Acknowledgements}
\noindent A.P.~and V.G.~are funded in part by the European Research Council (ERC) under the European Union's Horizon 2020 research and innovation programme under grant agreement No 757646 and UK STFC grant ST/P000630/1. F.\'O.h. is funded by a scholarship from the Scottish Funding Council.
A.J.~received funding from STFC consolidated grant ST/P000711/1 and from the European Research Council under the European Union's Seventh Framework Program (FP7/2007-2013) / ERC Grant agreement 279757. C.T.S is partially supported by an Emeritus Fellowship from the Leverhulme Trust. J.R acknowledges support from STFC for his studentship. This work used the DiRAC Extreme Scaling service at the University of Edinburgh, operated by the Edinburgh Parallel Computing Centre on behalf of the STFC DiRAC HPC Facility (www.dirac.ac.uk). This equipment was funded by BEIS capital funding via STFC capital grant ST/R00238X/1 and STFC DiRAC Operations grant ST/R001006/1. DiRAC is part of the National e-Infrastructure. The authors acknowledge the use of the IRIDIS High Performance Computing Facility in the completion of this work. 


\begin{thebibliography}{99}
{\footnotesize 
\bibitem{PhysRevD.91.074506}
N.~Carrasco, V.~Lubicz, G.~Martinelli, C.~T. Sachrajda, N.~Tantalo,
C.~Tarantino, and M.~Testa.
\newblock Qed corrections to hadronic processes in lattice qcd.
\newblock {\em Phys. Rev. D}, 91:074506, Apr 2015.

\bibitem{Foley:2005ac}
J.~Foley, K.~Jimmy Juge, A.~O'Cais, M.~Peardon, S.~M.~Ryan and J.~I.~Skullerud,
Comput.\ Phys.\ Commun.\  {\bf 172} (2005) 145
doi:10.1016/j.cpc.2005.06.008
[hep-lat/0505023].

\bibitem{Aoki2017}
S.~Aoki  \textit{et al}
\newblock Review of lattice results concerning low-energy particle physics.
\newblock {\em The European Physical Journal C}, 77(2):112, Feb 2017.

\bibitem{Hayakawa:2008an}
M.~Hayakawa and S.~Uno,
Prog.\ Theor.\ Phys.\  {\bf 120} (2008) 413
doi:10.1143/PTP.120.413
[arXiv:0804.2044 [hep-ph]].

\bibitem{Borsanyi:2014jba}
S.~Borsanyi {\it et al.},
Science {\bf 347} (2015) 1452
doi:10.1126/science.1257050
[arXiv:1406.4088 [hep-lat]].

\bibitem{Blum:2010ym}
T.~Blum, R.~Zhou, T.~Doi, M.~Hayakawa, T.~Izubuchi, S.~Uno and N.~Yamada,
Phys.\ Rev.\ D {\bf 82} (2010) 094508
doi:10.1103/PhysRevD.82.094508
[arXiv:1006.1311 [hep-lat]].

\bibitem{Davoudi:2018qpl}
Z.~Davoudi, J.~Harrison, A.~J{\"u}ttner, A.~Portelli and M.~J.~Savage,
arXiv:1810.05923 [hep-lat].

\bibitem{Boyle2017}
P.~Boyle, V.~G{\"u}lpers, J.~Harrison, A.~J{\"u}ttner, C.~Lehner, A.~Portelli,
and C.T. Sachrajda.
\newblock Isospin breaking corrections to meson masses and the hadronic vacuum
polarization: a comparative study.
\newblock {\em Journal of High Energy Physics}, 2017(9):153, Sep 2017.

\bibitem{Bloch:1937pw}
F.~Bloch and A.~Nordsieck,
Phys.\ Rev.\  {\bf 52} (1937) 54.
doi:10.1103/PhysRev.52.54

\bibitem{deDivitiis:2013xla}
G.~M.~de Divitiis {\it et al.} [RM123 Collaboration],
Phys.\ Rev.\ D {\bf 87} (2013) no.11,  114505
doi:10.1103/PhysRevD.87.114505
[arXiv:1303.4896 [hep-lat]].

\bibitem{Giusti:2017dmp}
D.~Giusti, V.~Lubicz, C.~Tarantino, G.~Martinelli, S.~Sanfilippo, S.~Simula and N.~Tantalo,
Phys.\ Rev.\ D {\bf 95} (2017) no.11,  114504
doi:10.1103/PhysRevD.95.114504
[arXiv:1704.06561 [hep-lat]].

\bibitem{Boyle:2008rh}
P.~A.~Boyle, A.~J\"uttner, C.~Kelly and R.~D.~Kenway,
JHEP {\bf 0808} (2008) 086
doi:10.1088/1126-6708/2008/08/086
[arXiv:0804.1501 [hep-lat]].

\bibitem{Allton:2008pn}
C.~Allton {\it et al.} [RBC-UKQCD Collaboration],
Phys.\ Rev.\ D {\bf 78} (2008) 114509
doi:10.1103/PhysRevD.78.114509
[arXiv:0804.0473 [hep-lat]].

\bibitem{lanczos}
C. Lanczos,
Journal of Research of the National Bureau of Standards
Vol.  45, No.4, October 1950, Research Paper 2133
}
\end{thebibliography}
\end{document}